\documentclass[prb,showpacs,twocolumn,superscriptaddress]{revtex4}
\usepackage{amsmath}
\usepackage{graphicx}
\usepackage{amsfonts}
\usepackage{amssymb}
\usepackage{hyperref}
\begin{document}
\title{Ultrafast (but Many-Body) Relaxation in a Low-Density Electron Glass}
\author{V.~K.~Thorsm{\o}lle}
\affiliation{\'Ecole Polytechnique F\'ed\'erale de Lausanne, CH-1015 Lausanne, Switzerland.}
\affiliation{D\'{e}partement de Physique de la Mati\`{e}re Condens\'{e}e, Universit\'{e} de Gen\`{e}ve, CH-1211 Gen\`{e}ve 4, Switzerland}
\author{N.~P.~Armitage}
\affiliation{Department of Physics and Astronomy, The Johns Hopkins University, Baltimore, MD 21218, USA.}
\date{\today}

\begin{abstract}
We present a study of the relaxation dynamics of the photoexcited conductivity of the impurity states in the low-density electronic glass, phosphorous-doped silicon Si:P.  Using subband gap optical pump-terahertz probe spectroscopy we find strongly temperature and fluence dependent glassy power-law relaxation occurring over sub-nanosecond time scales.   Such behavior is in contrast to the much longer time scales found in higher electron density glassy systems.  We also find evidence for both multi-particle relaxation mechanisms and/or coupling to electronic collective modes and a low temperature quantum relaxational regime.
\end{abstract}
\pacs{72.20.Ee, 71.30.+h, 71.45.Gm, 72.40.+w, 72.15.Rn} \vspace*{-10pt} \maketitle

Glasses are ubiquitous states of matter with positionally or rotationally randomized frozen degrees of freedom.  Their inherently frustrated interactions give rise to the existence of many energetically low-lying metastable states, which have a wide distribution of potential barriers separating them in configuration space.  At low temperatures, such systems are kinematically constrained from reaching their true ground-state on typical experimental time scales.  They are characterized by long relaxation times, memory, and aging effects \cite{Struik78a,Bouchaud98a,Vincent97a,BenChorin93a}.

In analogy with structural glasses, electronic glasses can be defined as systems with a random distribution of localized charges \cite{Davies82a,Mott79a,Pollak82a}.  Here long relaxation times and glassy phenomena derive from a combination of disorder and long-range unscreened Coulomb interaction.  The motion of any one charge manifestly necessitates a many-particle rearrangement of the other electron occupations to reach lower energy.  Such many-particle processes are inherently slow and inefficient at finding the true ground-state configuration.  Electronic glasses or `Coulomb glasses' may be realized in granular metals and amorphous and lightly doped semiconductors, which are all expected to exhibit certain similar qualitative behavior.  In addition to explicit glassy effects, there are also predictions for their equilibrium properties like the DC hopping conductivity \cite{Efros75a} and the power-law dependencies of the AC response \cite{Shklovskii81a,Lee01a,Helgren02a,Helgren04a}.

Recently there has been great progress in the understanding of how glassy non-equlibirum effects manifest themselves in these systems.  In amorphous compounds like In$_2$O$_{3-x}$ the natural history-free relaxation law of the DC conductivity after excitation is logarithmic over more than five decades in time \cite{Vaknin00a}, a behavior that may be rooted in a broad distribution of relaxation times.  Interesting temperature dependencies have been found at low temperature, which have been ascribed to quantum tunneling instead of thermally activated hopping \cite{Ovadyahu07a}.

The majority of such experiments have been on systems such as microcrystalline In$_2$O$_{3-x}$ or granular metals \cite{Vaknin00a,BenChorin93a,Grenet07a} in which the charge density is high.  It would be interesting to look for such phenomena in systems with much lower densities like doped semiconductors.  Although the equilibrium transport properties expected for electronic glasses have been reported in systems like Si:B or Si:P \cite{Lee01a,Helgren02a,Lee99a,Dai91a}, as far as we know, evidence for true glassy behavior in the conductivity, i.e. long-time non-exponential relaxation indicative of metastability and frustration has not been observed.  This is likely due to their much faster time scales owing to the much smaller charge densities, which is crucial for their quantum dynamics \cite{Ovadyahu07a}.  So while in principle glassy relaxation should exist in such systems, it is expected to be too fast to be resolved with usual techniques.

In this Letter we report the use of ultrafast optical pump-terahertz probe (OPTP) spectroscopy to resolve glassy relaxation in the doped semiconductor Si:P.  Terahertz time-domain spectroscopy is an ultrafast optical technique in which electric field transients are used to measure the conductivity of a material.  Here a sample is optically excited with a laser pulse and then probed at a later time with a terahertz (THz) pulse to measure the induced conductivity changes with picosecond resolution.  While OPTP has been successfully applied to correlated electron systems such as high-Tc's, manganites and semiconductors  \cite{Averitt02a,Jepsen01a} in unraveling the various dynamics by their different relaxation time scales, this is the first such study on glasses.  In this work, we find strongly temperature and fluence dependent glassy non-exponential relaxation occurring on a sub-nanosecond time scale, which is in contrast to the much longer time scales observed in high electron density glassy systems.

Experiments were performed on nominally uncompensated phosphorous-doped silicon samples, which  were cut from a Czochralski grown boule grown by Recticon Enterprises Inc.~to a specification of 5~cm in diameter with a P-dopant gradient along the axis. This boule was subsequently sliced and then polished down to 100~$\mu$m.  Samples from this boule have previously been used for an extensive study of the THz-range conductivity in the phononless regime \cite{Helgren02a,Helgren04a}.  In this study, we concentrated on a particular Si:P sample with a 300~K resistivity of 0.0202~$\Omega \cdot$cm.  According to the Thurber scale \cite{Thurber80a} this corresponds to an estimated phosphorus concentration $n$ of $1.37\times10^{18}$ dopant atoms/cm$^{3}$, which puts it at 39$\%$ of the way to the MIT.  This sample's  \cite{Helgren04a} localization length ($\approx$ 13 nm) is slightly larger than the interdopant spacing of $\approx$ 9 nm and its static dielectric constant ($\approx$ 14) is enhanced over that of undoped silicon (11.7) demonstrating that collective quantum effects may play a role.  Although nominally uncompensated, it is also generally believed that this doping level is high enough that the material self-compensates i.e. the long-range Coulomb interaction ionizes a substantial fraction of P sites.

\begin{figure}[htbp]
\begin{center}
\includegraphics[width=7.5cm,angle=0]{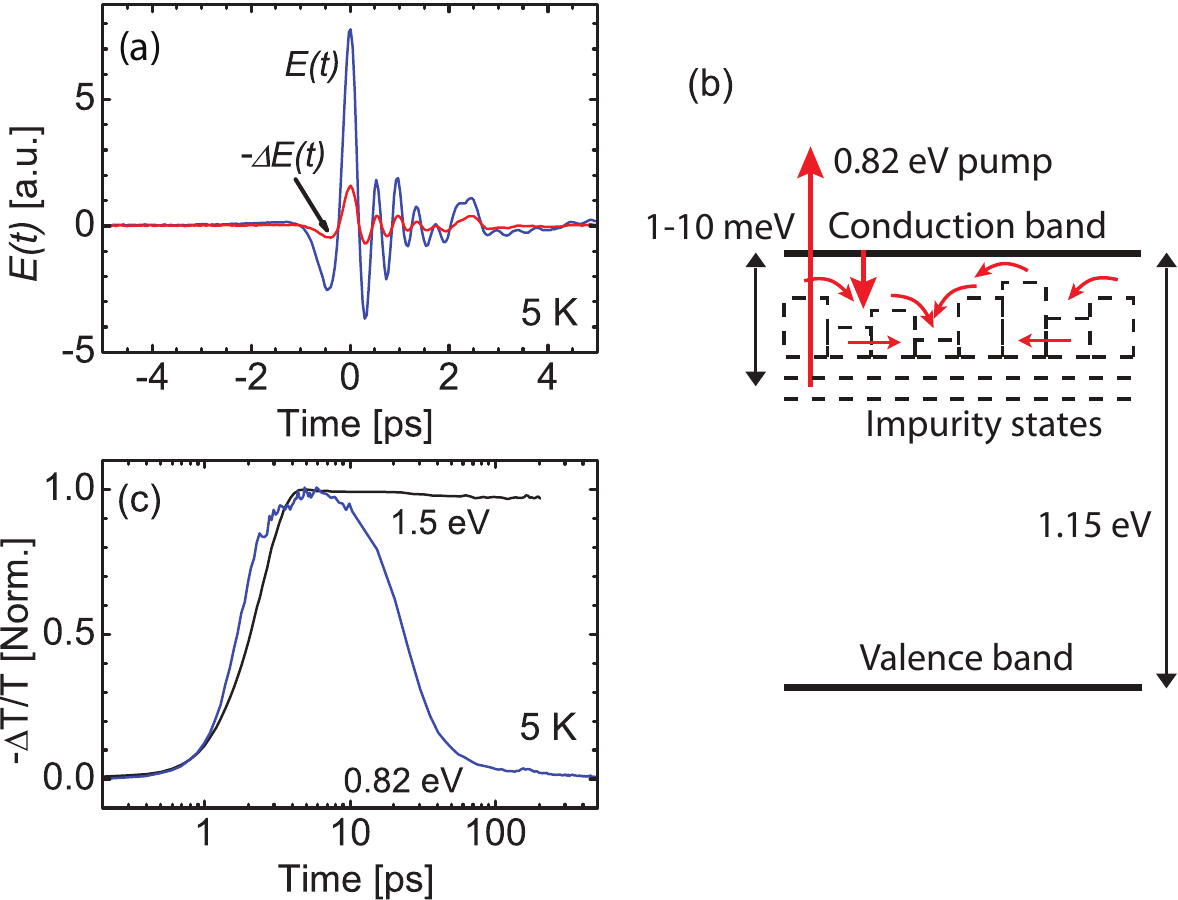}
\caption{(color) (a) Electric field $E(t)$ of transmitted THz pulse, and induced change in electric field $-\Delta E(t)$ at $T=5$~K.  (b) Schematic energy diagram of Si:P with impurity states close to the conduction band. The upward arrow indicates photoexcitation at 0.82~eV. The downward arrows indicate the subsequent relaxation processes.  (c) Comparison of the normalized differential transmission of the THz probe pulse peak $\Delta\mbox{T}/\mbox{T}$ versus delay with a pump excitation energy above (1.5~eV) and below (0.82~eV) the band gap at $T=5$~K for a photoexcited carrier density of $3 \times 10^{17}$ cm$^{-3}$ in both cases.}
\label{Fig1}
\end{center}
\end{figure}

The experiments utilized a regeneratively amplified Ti:Al$_2$O$_3$ laser system operating at 1 kHz, producing nominally 1.0~mJ, 150~fs pulses at 1.5~eV. The Si:P sample was excited at 1500~nm (0.82~eV) $via$ an optical parametric amplifier with pump fluences ranging from 4.2 to 424.4~$\mu$J/cm$^2$ producing photoexcited carrier densities \cite{Densities} $n_{\mbox{pe}}\sim3.2\times10^{15}-3.2\times10^{17}$~cm$^{-3}$.  The experiments were performed in transmission (T) with the Si:P sample inside an optical He cryostat capable of reaching 4~K.  The THz `probe' pulses were generated and detected using electro-optic techniques.  A schematic of the THz setup is given in Ref. \onlinecite{Thor07a}.  Fig. \ref{Fig1}a shows the transmitted electric field of the THz pulse before $E(t)$ and after $\Delta E(t)$ optical excitation. The decrease in the transmitted electric field is associated with the change in conductivity (i.e., $\Delta\sigma\varpropto -\Delta T/T$).   From an analysis of the skin-depth, the sample is presumed to be homogeneously excited at low temperature.  In equilibrium, the THz range conductivity in these materials derives from transitions between localized levels and is believed to be proportional to a high power of the localization length $\xi$ as $\sigma_1 \propto N_0^2 \xi^4$ where $N_0$ is the non-interacting density of states  \cite{Shklovskii81a,Lee01a,Helgren02a,Helgren04a}.

The excitation process is shown in Fig.~\ref{Fig1}b.  Electrons are excited from the impurity band to a relatively large energy in the conduction band  (0.8 eV).  However, they quickly decay back into impurity states d through successive many-body hops find lower energy orbitals, slowly reducing the transmission.  In order to probe only impurity band states it is essential that the pump laser's energy is below the band gap threshold (1.5 eV).  Note that the dynamics of conduction band-valence band recombination is completely different as shown in Fig.~\ref{Fig1}c.  Recombination of 1.5 eV electron-hole pairs produced by above-gap excitation occur on time scales of millisecond or longer, while 0.8 eV excitations decay on sub-nanosecond time scales.   While recombination of conduction band - valence band pairs from a 1.5 eV excitation pulse is slow it can be described by a simple exponential, and is $not$ glassy.

\begin{figure}[tbp]
\begin{center}
\includegraphics[width=7cm,angle=0]{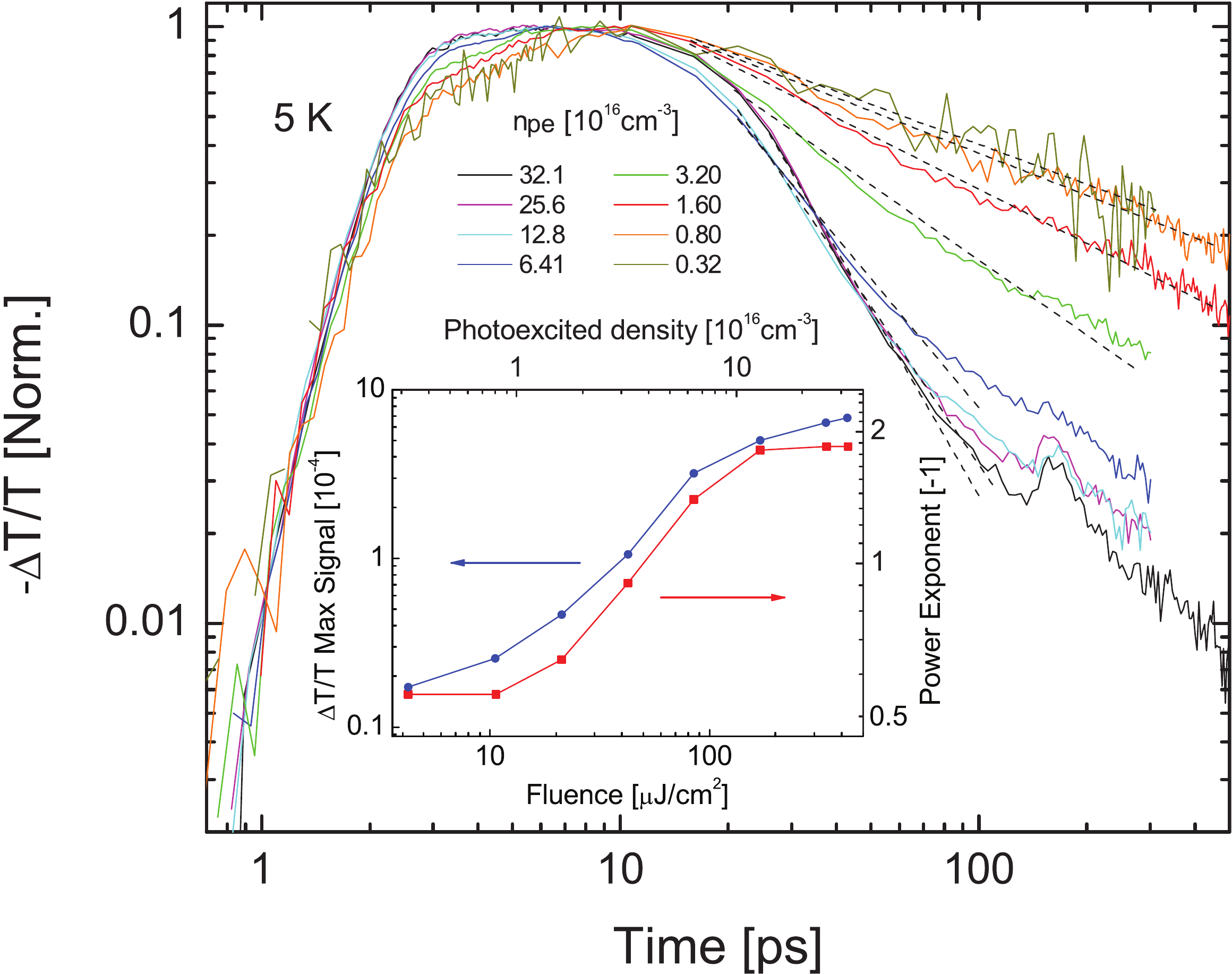}
\caption{(color) Time-resolved $\Delta\mbox{T}/\mbox{T}$ photoinduced transmission dynamics of Si:P sample at different $n_{\mbox{pe}}$'s at $T=5$~K and an excitation energy of $0.82$~eV. The data are normalized to maximum signal. Fits are shown as black dashed lines. The inset shows the photoexcited charge density $n_{\mbox{pe}}$ and fluence dependence of the maximum $\Delta\mbox{T}/\mbox{T}$ signal together with the exponent of a power-law for the relaxation.}
\label{Fig2}
\end{center}
\end{figure}

In Fig.~\ref{Fig2} we show the peak normalized photoinduced $\Delta\mbox{T}/\mbox{T}$ dynamics at various $n_{\mbox{pe}}$'s at $T=5$~K with 0.8 eV excitation.  All the data present a fast $\sim$1 picosecond rise, a plateau, and then a much slower decay.   As we will discuss below, the data have a number of unusual features consistent with glassy dynamics.  At high $n_{\mbox{pe}}$ the decay is power-law  with an exponent of $\sim$-2 until $\sim$180 ps.  At $n_{\mbox{pe}}$ below $10^{17}$~cm$^{-3}$, the relaxation is still power-law (Fig.~\ref{Fig2} inset) with an exponent leveling off at $\sim$-0.5 below $n_{\mbox{pe}}\simeq1.5\times10^{16}$~cm$^{-3}$ that persists out to at least $\sim$500 picosecond \cite{stretched}.

The plateau region in $\Delta\mbox{T}/\mbox{T}$ is associated with the decay of charge into the impurity band into quasi-equilbrium.  All subsequent relaxation concerns the rearrangement of charge within this band of localized states.  Relaxation occurs within localized states as the form of the THz conductivity does not change after photoexcitation, as shown by the essentially identical shape of E-field transients in Fig. \ref{Fig1}a.  Therefore $\sigma$$\rightarrow$0 as $\omega$$\rightarrow$0 just as in the equilibrium case and as one expects for AC transport within localized states.  Therefore $\sigma$$\rightarrow$0 as $\omega$$\rightarrow$0 just as in the equilibrium case and as one expects for AC transport within localized states.  Note also that despite excited charge densities of at most 23$\%$ of $n$, the overall scale of the transmission change (Fig. \ref{Fig2} inset) never reaches more than 0.1$\%$ of the unpumped transmission even at the highest $n_{\mbox{pe}}$.   This gives additional evidence that the decay should be associated with relaxation within a localized band of low intrinsic conductivity.

The strong $n_{\mbox{pe}}$ dependence of the instantaneous decay rate and power-law is unusual and evidence for multi-particle relaxation.   Still, even in many systems where two-particle recombination kinetics are important, the Rothwarf-Taylor ``phonon-bottleneck'' effect typically sets in, which decreases the $n_{\mbox{pe}}$ dependence \cite{Rothwarf67a}.   In such a case, the phonons created $via$ electron-hole pair decay are available to make new electron-hole pairs, which limits the net energy flow out of the electronic system and diminishes the role of two-particle reactions as a rate limiting step.  Effects of this kind apparently are not dominant here.  The $n_{\mbox{pe}}$ dependence is also different than relaxation from above gap excitations in other disordered semiconductors like microcrystal silicon \cite{Jepsen01a} where the decay rate $decreases$ with increasing $n_{\mbox{pe}}$.   There, with increasing $n_{\mbox{pe}}$, the available trapping sites are filled up and only slow conduction band - valence band recombination is available as a decay channel.  Obviously, the functional dependencies here are very different.

We propose that the power-law decay and the strong tendency towards a smaller exponent at low $n_{\mbox{pe}}$ is consistent with glassy many-body relaxation.   After initial photoexcitation charges fall back to the impurity band into a quasi-equilibrium configuration, which necessarily occupies higher energy levels where localization is weaker.  The system is constrained through large potential barriers from reaching lower energy configurations expeditiously.  It is only through rare and intrinsically slow many-particle rearrangements that electrons find lower energy configurations thereby reducing the conductivity.   These relaxations may occur through Auger-like rearrangements  \cite{Landsberg87a} of a discrete number of electrons or through a coupling to collective electronic modes \cite{Mueller08a}.

Although we find power-law and not logarithmic decay as in the DC experiments on In$_2$O$_3$  \cite{Vaknin00a}, note that power-law $t^\alpha$ decay with $\alpha \ll 1$ has a dependence very close to logarithmic.  Like the logarithmic dependence the power-law presumably derives also from an average over a broad distribution of relaxation times.  Although a power-law decay is by definition scale invariant, it is in the sense of the small power-law exponent that the relaxation can be considered `slow' and glassy.

In the present case, despite the decay's scale-free nature we can compare the rough time scales in our experimental range to other natural relaxational scales.  At 5~K and $n_{\mbox{pe}}\simeq3.2\times10^{15}$~cm$^{-3}$, it takes $\approx$200~ps for $\Delta T / T$ to decay to 30$\%$ of its maximum.   This is much longer than for instance, the natural scale for phonon assisted relaxation, which should be of order $\frac{1}{\omega_D} e^{  ({\frac{T_{ES}}{T}})^{1/2}} \approx $ 9 picosecond in this sample at 5 K \cite{Mott79a}.  Here $\omega_D$ is the Debye frequency of silicon and $T_{ES}$ is the characteristic Efros-Shklovskii temperature $\frac{1}{k_B}\frac{e^2}{4 \pi \epsilon \xi}$.  It is also interesting to compare to the `Maxwell time' $\tau_M$, which is the classical expectation for spatial charge relaxation \cite{BenChorin93a}.  $\tau_M$ calculated from the 5 K DC resistivity\cite{Helgren04a} $\rho_{DC}$ is approximately 0.9 millisecond ($\tau_M = \rho_{DC}/4 \pi \epsilon_1 \epsilon_0$), which is far in excess of any scale we observe.   However, it may be possible to define an `AC Maxwell time' from the equilibrium THz range resistivity associated with short length scale relaxation.   Such a time calculated from older data at our central THz frequency (600 GHz) \cite{Helgren04a} is 14 ps.  The difference between these time scales and the experimental ones may also be evidence for glassy effects.

\begin{figure}[tbp]
\begin{center}
\includegraphics[width=7cm,angle=0]{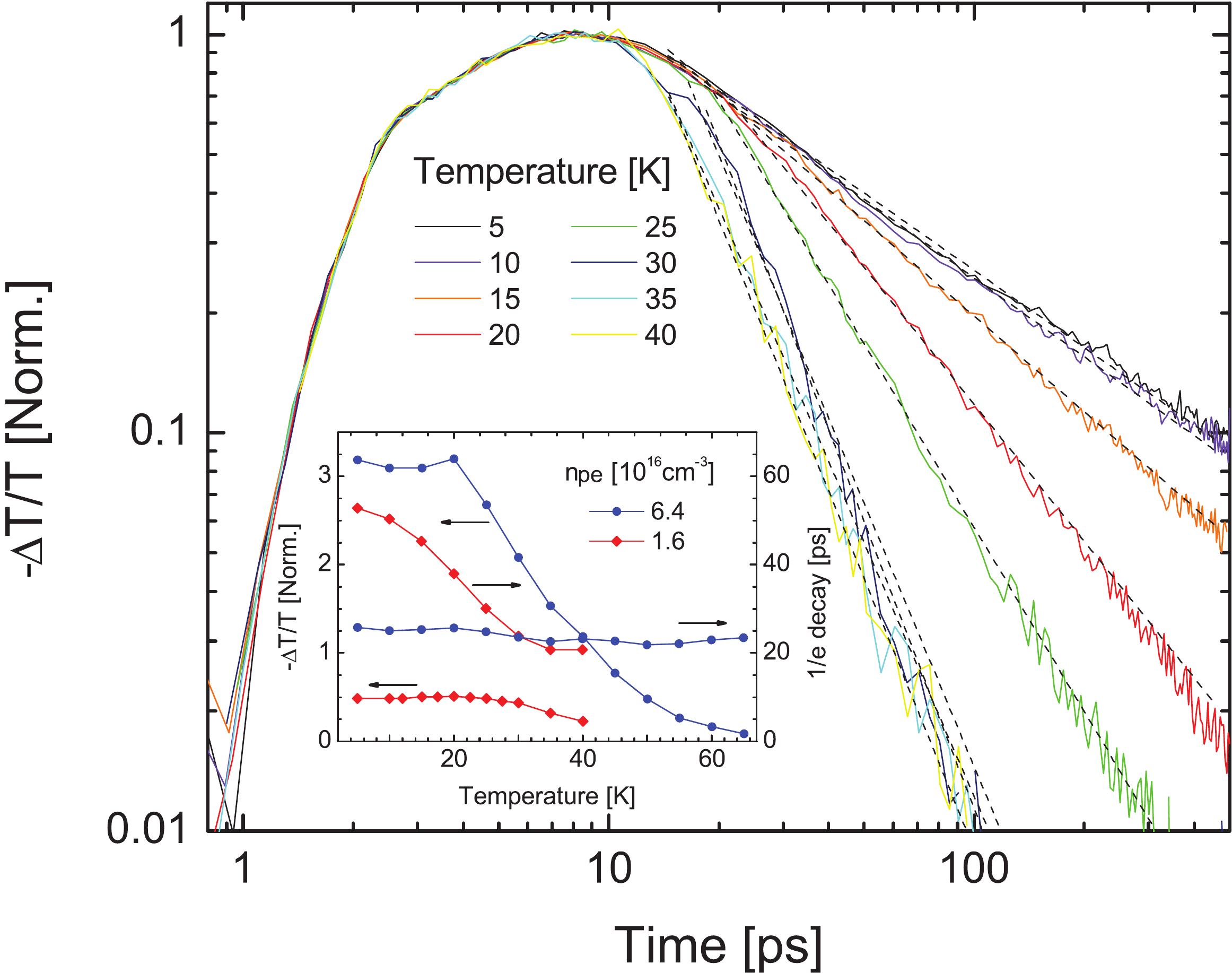}
\caption{(color) Time-resolved $\Delta\mbox{T}/\mbox{T}$ photoinduced transmission dynamics of Si:P sample at different temperatures for $n_{\mbox{pe}}\simeq1.6\times10^{16}$~cm$^{-3}$ and an excitation energy of $0.82$~eV. The data are normalized to maximum signal. Power-law fits are shown as black dashed lines. The inset shows the temperature dependence of the maximum $\Delta\mbox{T}/\mbox{T}$ signal together with the time it takes to decay to 1/e for $n_{\mbox{pe}}\simeq1.6\times10^{16}$~cm$^{-3}$ and $n_{\mbox{pe}}\simeq6.4\times10^{16}$~cm$^{-3}$. }
\label{Fig3}
\end{center}
\end{figure}

\begin{figure}[t]
\begin{center}
\includegraphics[width=7.5cm,angle=0]{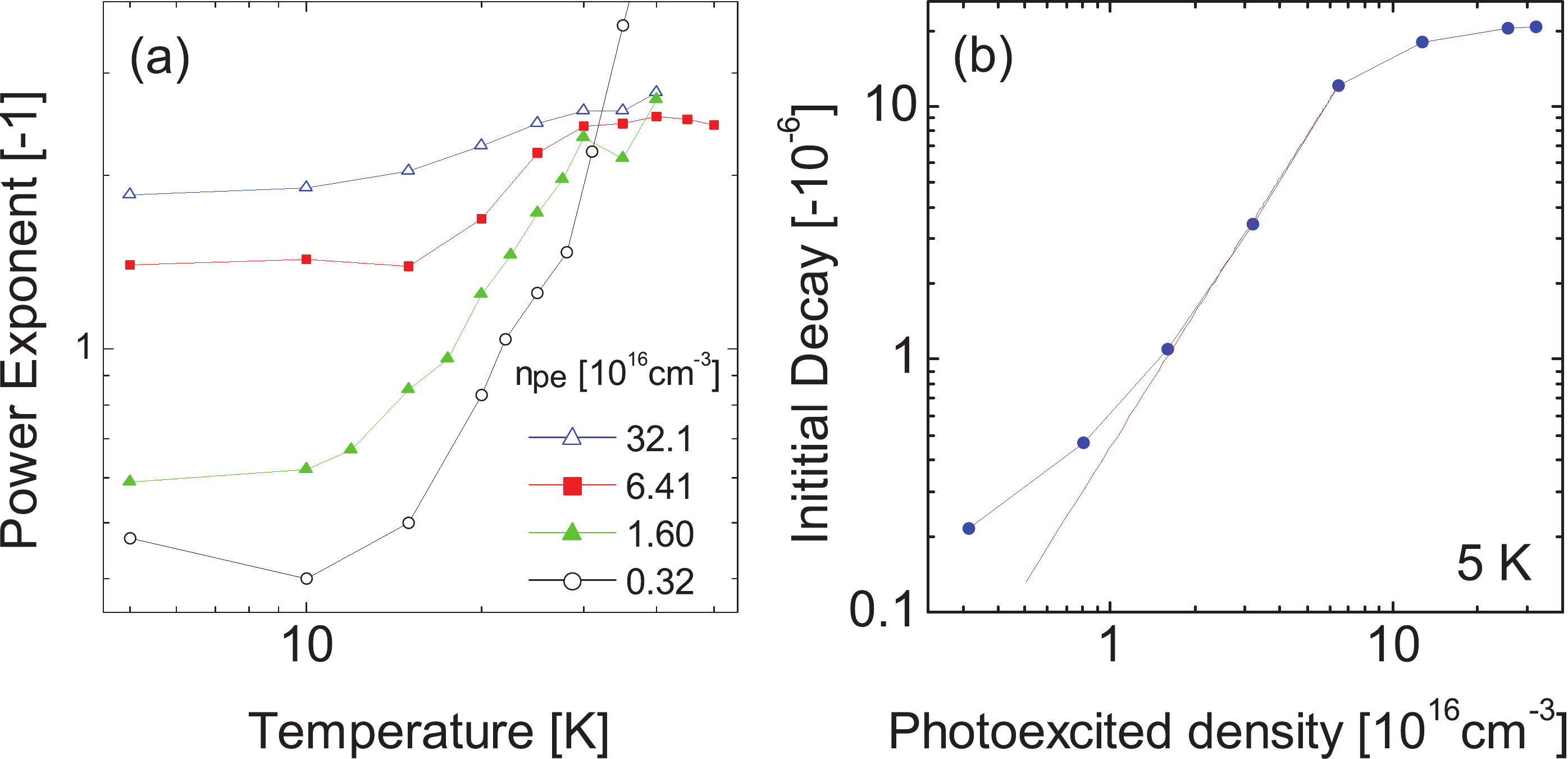}
\caption{(color) (a) Exponent of power-law decay vs. temperature at different photoexcited carrier densities.   A low temperature saturation region is seen which is temperature independent below some $n_{\mbox{pe}}$.  (b)   Initial decay rate after initial excitation vs. $n_{\mbox{pe}}$.   The red curve is a power-law fit to the intermediate $n_{\mbox{pe}}$ region with exponent $\approx 1.8$. }
\label{Fig4}
\end{center}
\end{figure}

The temperature dependence supports our claim of glassy relaxation.  Fig. \ref{Fig3} shows the normalized time-resolved photoinduced $\Delta\mbox{T}/\mbox{T}$ at various temperatures for $n_{\mbox{pe}}\simeq1.6\times10^{16}$~cm$^{-3}$. It features a $\sim$1 picosecond rise, followed by a nearly constant signal over $\sim$7 picosecond after which it displays a temperature dependent power-law decay.  At the lowest temperature the power-law exponent for this $n_{\mbox{pe}}$ is $\sim$-0.6, and becomes about $\sim$-2.5 above 25~K. The inset shows the dependence of the transmission maximum for both $n_{\mbox{pe}}\simeq1.6\times10^{16}$~cm$^{-3}$ and $n_{\mbox{pe}}\simeq6.4\times10^{16}$~cm$^{-3}$.  In Fig. \ref{Fig4}a we show a compilation of the temperature dependent fits of the power-law decay at different $n_{\mbox{pe}}$. Above a threshold temperature of around $\sim$25 K and  $n\simeq1.6\times10^{16}$~cm$^{-3}$, the dynamics shows no significant changes with $n_{\mbox{pe}}$ as is expected for such a system out of the glassy regime. Below $\sim$25~K, the power-law relaxation becomes slower.  As noted previously the very lowest measured $n_{\mbox{pe}}$ exhibits a power-law very close to 0.5 at the lowest temperature.  Note that all measured curves show a low temperature region where the exponent saturates.   This region shrinks slightly with decreasing temperature, but remains finite even in the limit of low $n_{\mbox{pe}}$.   This is consistent with quantum tunneling as means to relaxation.  In contrast, if the relaxation was purely due to thermal effects or dependent on the energy deposited from the pump pulse, one would expect that the temperature threshold for saturation and the exponent to go to zero continuously at low T and $n_{\mbox{pe}}$.

We have shown evidence for a highly temperature and $n_{\mbox{pe}}$ dependent relaxation consistent with glassy dynamics.   What can we say more precisely about the nature of the relaxation?  In materials where two-particle recombination is the principle relaxation channel, one expects a $ - \beta n_{\mbox{pe}}^2$ term in the rate equations and an instantaneous decay rate that goes as $\beta n_{\mbox{pe}}$.  This is what has been found in systems like some superconductors where electron-hole recombination dominates \cite{Averitt02a,Gedik04a}.  In glassy systems, simultaneous multi-particle hoppings should contribute terms to the rate equations as a high order in $n_{\mbox{pe}}$ and hence the instantaneous rates would be proportional to $n_{\mbox{pe}}$ to a power larger than unity.   One may also expect a super-linear dependence if relaxation is driven by excited collective electronic modes \cite{Mueller08a}.

In Fig. \ref{Fig4}b we plot the instantaneous decay rate from a 10 picosecond time span after the transmission plateau vs. $n_{\mbox{pe}}$.   Strictly exponential decay would be linear and two-particle relaxation quadratic on this plot.  The data shows a dependence faster than linear and at intermediate $n_{\mbox{pe}}$'s goes like the $\approx 1.8$ power.  This is consistent with multi-particle relaxation.  Due to the previously mentioned temperature and $n_{\mbox{pe}}$ independence at low $n_{\mbox{pe}}$, for a full description it seems essential to incorporate quantum relaxation effects.   These are not normally considered in relaxation models and theoretical input is essential for a more rigorous interpretation.

We thank J. Demsar, E. Helgren, D. Hilton, M. M\"uller, M. Pollak, and Z. Ovadyahu for helpful discussions and/or careful reading of this manuscript.   The research at JHU was supported by NSF DMR-0847652.

\bibliography{EGlassBib}

\begin{thebibliography}{27}
\bibitem[{Den()}]{Densities}
\bibinfo{note}{$n_{\mbox{pe}}$ was obtained using $n_{\mbox{pe}}\simeq
  I\lambda/hcd$, where $I$ is fluence, $\lambda$ excitation wavelength, and $d$
  thickness.}
\bibitem[{str()}]{stretched}
\bibinfo{note}{In practice it is quite difficult to exclude functional forms
  such as stretched exponential, but we were only able to fit stretched
  exponential to the lowest $n_{\mbox{pe}}$ and T data.}
\end{thebibliography}

\end{document}